\RequirePackage{fix-cm}
\documentclass[smallextended,envcountsect]{article}       
\usepackage{amsfonts}
\usepackage{amsmath}
\usepackage{color}
\usepackage{cleveref}
\usepackage{bm}
\usepackage{stmaryrd}
\usepackage{algorithm}
\usepackage{algpseudocode}
\usepackage[utf8]{inputenc}
\usepackage[english]{babel}
\usepackage{graphicx}
\usepackage{enumerate}
\usepackage[misc]{ifsym}
\usepackage{subcaption}
\usepackage[style=base]{caption}
\usepackage{ulem}
\usepackage{cancel}
\usepackage{lineno}
\usepackage{url}

\usepackage[authoryear]{natbib}
\bibpunct{(}{)}{}{a}{}{;}

\begin{document}

\title{A comparison between Bayesian and ordinary kriging based on validation criteria - Application to radiological characterisation
}

\author{Martin Wieskotten\\
CEA, DES, ISEC, DMRC, Univ. Montpellier, Marcoule, France \\
LMA Université d'Avignon, EA 2151, 84029, Avignon, France \\
\\
Marielle Crozet\\
CEA, DES, ISEC, DMRC, Univ. Montpellier, Marcoule, France \\
\\
Bertrand Iooss\\
EDF R\&D, 6 quai Watier, 78400, Chatou, France \\
Institut de Mathématiques de Toulouse, France \\
\\
Céline Lacaux  \\
LMA Université d'Avignon, EA 2151, 84029, Avignon, France \\
\\
Amandine Marrel\\
CEA, DES, IRESNE, DER, Cadarache, Saint-Paul-Lez-Durance, France \\
Institut de Mathématiques de Toulouse, France 
}

\maketitle

\begin{abstract}
In decommissioning projects of nuclear facilities, the radiological characterisation step aims to estimate the quantity and spatial distribution of different radionuclides.
To carry out the estimation, measurements are performed on site to obtain preliminary information.
The usual industrial practice consists in applying spatial interpolation tools (as the ordinary kriging method) on these data  to predict the value of interest for the contamination (radionuclide concentration, radioactivity, etc.) at unobserved positions.
This paper questions the ordinary kriging tool on the well-known problem of the overoptimistic prediction variances due to not taking into account uncertainties on the estimation of the kriging parameters (variance and range).
To overcome this issue, the practical use of the Bayesian kriging method, where the model parameters are considered as random variables, is deepened.
The usefulness of Bayesian kriging, whilst comparing its performance to that of ordinary kriging, is demonstrated in the small data context (which is often the case in decommissioning projects).
This result is obtained via several numerical tests on different toy models, and using complementary validation criteria: the predictivity coefficient ($Q^2$), the Predictive Variance Adequacy ($PVA$), the $\alpha$-Confidence Interval plot (and its associated Mean Squared Error $\alpha$ ($MSE\alpha$)), and the Predictive Interval Adequacy ($PIA$).
The latter is a new criterion adapted to the Bayesian kriging results.
Finally, the same comparison is performed on a real dataset coming from the decommissioning project of the CEA Marcoule G3 reactor.
It illustrates the practical interest of Bayesian kriging in industrial radiological characterisation.

Keywords: Geostatistics, Bayesian kriging, Ordinary kriging, Validation criteria, Radiological characterisation more
\end{abstract}

\section{Introduction}
\label{sec:1}

Radiological characterisation is one of the main challenges encountered in the nuclear industry for the decommissioning and dismantling (D\&D) of old infrastructures such as buildings (see, e.g., \citet{cet14}, \citet{epri} and \citet{cea_den_assainissement-demantelement_2017}). Its main goal is to evaluate the quantity and spatial distribution of radionuclides. As such, measurements are made to constitute a dataset and obtain preliminary information. While measurements are made, many problems can arise. The radioactivity present on site can be dangerous for operators and does not allow for many measurements. In some extreme cases, drones and robots have to be used, making measurements more expensive and reducing the size of the datasets  (see, e.g., \citet{gougal15} and \citet{cea_den_assainissement-demantelement_2017}). It is therefore quite common in nuclear D\&D characterisation to have only a small number of available data: a balance has to be found between data acquisition costs and provided information from data.
Statistical tools make it possible to optimise the information extracted from the data, within a rigorous mathematical framework and with associate confidence intervals (in the D\&D field, see, e.g., \citet{zafmag16}, \citet{bladel17} and \citet{perlec20}).

More precisely, as in many other environmental and industrial fields (see, e.g., \citet{webster_geostatistics_2007} and \citet{sagar2018handbook}), spatial statistics and geostatistical methods are used to predict the variables of interest at an unobserved location (prediction of the expected value), with an indication of the expected error in prediction (prediction variance). The methodology is often based on two steps: first the construction of a statistical model with the estimation of its parameters, followed by the prediction with the statistical model for any unobserved point. 
The ordinary kriging model (see, e.g., \citet{chiles_geostatistics_2012} and \citet{cressie_statistics_1993}) is one of the most widely used models in industrial practice of D\&D (see, e.g., \citet{cet14}, \citet{gougal15} and \citet{epri}). 
However, a common criticism is that its predictions do not take into account the uncertainty in the estimation of the model parameters.
As a result, the variances of the predictions are often too optimistic and these neglected uncertainties in the model parameters can have a significant impact.
This problem is made worse for smaller datasets, which can be common in D\&D projects. 
For the radiological characterisation in D\&D projects, the first examples of kriging shown in \citet{jeades08}, \citet{desnoyers_approche_2010} and \citet{deschi11} have studied practical cases based on many measurements and did not consider this issue.
The more realistic studies by \citet{bodrog13}, \citet{lajren20} and \citet{desfau20}, carried out on smaller datasets, have instead highlighted the errors generated by the estimation errors of the kriging parameters.

To overcome this kriging issue, a Bayesian approach was first proposed by \citet{kitanidis_parameter_1986}. Its main goal was to take into account the uncertainties in the scale and mean parameters of the kriging model. The work of \citet{handcock_bayesian_1993} then completed the full Bayesian approach which considers all the parameters of the model as unknown. More recently, a slightly different approach was presented by \citet{krivoruchko_evaluation_2019} and is called empirical Bayesian kriging. This methodology differs slightly from the one used in \citet{kitanidis_parameter_1986}, since the choice on the prior distributions of kriging parameters are obtained through unconstrained simulations of the random field. This approach was adapted to allow for multi-fidelity applications, where Bayesian theory is used to update the initial data with new, more accurate data (classically used with cokriging if correlations between old and new data exist). Some examples can be found in meteorology in \citet{gupta_comparison_2017} or for oil extraction in \citet{al-mudhafar_bayesian_2019}. Note that a more complete description of Bayesian kriging with an extension to generalised linear models is presented in \citet{diggle_model-based_2007}.

In this framework, our work aims to understand the usefulness of the Bayesian kriging approach, compared to the ordinary kriging one, for the radiological characterisation of contaminated buildings.
In particular, the specification of a priori laws for the parameters in Bayesian kriging, which allows a more robust estimation of these parameters when only a few observations are available, is studied.
The performance of ordinary and Bayesian kriging is compared on several numerical examples. 
For this, we not only focus on the kriging predictor accuracy but also on the kriging predictive variance accuracy.
Indeed, the kriging variance is often used by practitioners to estimate predictive intervals on predicted quantities, to justify their choice of sampling, or to find locations of new (potentially expensive) measurements \citep{becrom13}.
To ensure a certain level of confidence in the use of the predictive variance, the works of \citet{marioo12}, \citet{bac13}, \citet{demay_model_2022} and \citet{acharki2023}, about kriging model validation, have emphasised the usefulness of several validation criteria, as the Predictive Variance Adequacy ($PVA$) and the $\alpha$-Confidence Interval ($\alpha$-CI) plot.
In addition to allow a more accurate comparison in the case of the Bayesian kriging model, new validation criteria are required and are proposed in the present work.

The following section describes the different studied kriging models, while Sect. 3 develops the associate classical validation criteria before introducing the newly proposed ones. 
Section 4 presents the results of the model comparison obtained on several numerical tests.
Section 5 then illustrates the application on a real case study coming from the decommissioning project of the CEA Marcoule G3 reactor. 
Section 6 gives some conclusions.
Finally, two appendices present prior specification and parameter estimation results, which are not discussed in the main work of this article.

\section{The ordinary and Bayesian kriging models}
\label{sec:2}

This section provides reminders on kriging principles, within the framework of Gaussian random field model.

\subsection{The Gaussian random field model}
\label{sec:2.1}

The variable of interest is assumed to be a random field $\{Z(\boldsymbol{x}), \boldsymbol{x}\in D\}$, with $D\subset\mathbb{R}^{2}$. $Z(.)$ is supposed to be isotropic and stationary, meaning that
$$\forall \boldsymbol{x} \in D, E[Z(\boldsymbol{x})]=\beta,$$
$$\forall \boldsymbol{x},\boldsymbol{x'} \in D,\mbox{Cov}(Z(\boldsymbol{x}),Z(\boldsymbol{x'}))=\sigma^2 C_{\phi} (|\boldsymbol{x}-\boldsymbol{x'}|),$$
where $C_{\phi}$ is the correlation function where $C_{\phi}(0)=1$, and $\beta, \sigma^2, \phi$ denote the mean, variance and range (or correlation length) parameters, respectively. For ease of notation, the conditioning to parameters will be simplified from $Z|\beta =\widehat{\beta}$  to $Z|\beta$. The term $C_{\phi}$ corresponds to a positive semi-definite function.
Moreover, by definition of a Gaussian process, every finite set of $Z$ is a multivariate normal distribution (denoted ${\cal N}(.,.)$). Thus for $n$ observations at positions ${\boldsymbol{x}_1,\ldots,\boldsymbol{x}_n}$, we obtain the Gaussian random vector $Z=(Z(\boldsymbol{x}_1),\ldots,Z(\boldsymbol{x}_n))'$ with
$$Z|\beta,\sigma^2,\phi \sim {\cal N}(\beta \boldsymbol{1}_n,\sigma^{2}\boldsymbol{R}_{\phi}),$$
where $\boldsymbol{1}_n$ is the vector $(1,\ldots,1)'$ of length $n$, and the covariance matrix is $\sigma^{2}\boldsymbol{R}_{\phi}~=~\left(\mbox{Cov}(Z(\boldsymbol{x}_i),Z(\boldsymbol{x}_j))\right)_{1\le i,j \le n}$.
The observation sample of $Z$ is written $\boldsymbol{z}~=~(z(\boldsymbol{x}_1),\ldots,z(\boldsymbol{x}_n))'$.

The positive semi-definite function $C_{\phi}$ is often modeled using common covariance function. In this work, two covariance models will be used (see, e.g., \citet{chiles_geostatistics_2012} for an extensive list of covariance functions).
The first one is the Gaussian covariance function written
$$\forall h \in \mathbb{R}, C_{\phi}(h)=e^{-h^2/\phi^2},$$
while the second one is the Matérn covariance function written
\begin{equation}\label{eq:matern}
\forall h \in \mathbb{R}, C_{\phi,\nu}(h)=\frac{2^{1-\nu}}{\Gamma(\nu)}\left(\sqrt{2\nu}\frac{h}{\phi}\right)^\nu K_{\nu}\left( \sqrt{2\nu}\frac{h}{\phi}\right),
\end{equation}
with $\nu$ a strictly positive parameter, $\Gamma(.)$ the gamma function and $K_\nu(.)$ the modified Bessel function of second type and order $\nu$. The parameter $\nu$, that drives the regularity of the process trajectories, is not estimated. 
It is chosen from a set of possible values, the most commonly used being $\nu \in \{ \frac{1}{2},\frac{3}{2},\frac{5}{2}\}$.
In addition, we have the nugget effect, written
$$\forall h \in \mathbb{R}, C_{\tau^2}(h)=\tau^2\delta(h),$$
with $\tau^2$ a variance and $\delta$ the Dirac function where $\delta(h)=1$ if $h=0$ and $\delta(h)~=~0$ otherwise. The nugget effect is often used to model micro-scale variations and measurements uncertainties. In our case studies, it will mainly be used to improve the conditioning of the matrix $\boldsymbol{R}_{\phi}$, in order to improve the stability of its numerical inversion (especially in the case of Gaussian covariance function).

The model is therefore specified by three different parameters: the trend parameter $\beta \in D_\beta$, the variance parameter $\sigma^2 \in D_{\sigma^2}$ and the range parameter $\phi \in D_{\phi}$. In the case of ordinary kriging and for the covariance functions considered here, the parameter spaces are
$$D_\beta = \mathbb{R}, D_{\sigma^2} = ]0,+\infty[, D_{\phi} = ]0,+\infty[.$$

The first step of the kriging methodology in practice is to estimate these parameters. Two main procedures are commonly used: variographic analysis and maximum likelihood estimation (MLE). An extensive literature is available about parameter estimation with variographic analysis, such as \citet{chiles_geostatistics_2012} and \citet{webster_geostatistics_2007}. In this work, we will use maximum likelihood estimation to take advantage of the probabilistic framework and to avoid manual or automatic fitting of variograms, especially since our numerical tests will require parameter estimation for many simulated datasets.
Moreover, the automatic fitting of variograms is strongly discouraged in most of the literature (see, e.g., \citet{chiles_geostatistics_2012} and \citet{webster_geostatistics_2007}). 
Note that, when kriging is used to interpolate and predict numerical experiments with a large number of inputs, a multi-start optimization procedure is often used for the MLE to avoid the known pitfall of local extrema and better explore the input parameter space. However, this procedure will not be used here because preliminary studies have shown that in our case it is not necessary due to the small dimension of the problem (2D, i.e., two-dimensional random field) and the regularity of the likelihood function. This decision allowed to reduce computation times without compromising on parameter estimation.

\subsection{Kriging model principles}
\label{sec:2.2}

The kriging predictor is a linear interpolator whose expressions are derived from supplementary conditions, such as minimizing the prediction variance. For a detailed description of kriging and its construction, the reader can refer to the reference books of \citet{chiles_geostatistics_2012}, \citet{cressie_statistics_1993} for geostatistics, but also \citet{rasmussen_gaussian_2006} for the Gaussian process regression point of view. Let $\boldsymbol{x}_0$ be an unobserved position at which we wish to predict the expected value and the variance of $Z(\boldsymbol{x}_0)|\sigma^2 ,\phi,\boldsymbol{Z}=\boldsymbol{z}$ (the mean is considered unknown).
The ordinary kriging equations are then
$$\mathbb{E}[Z(\boldsymbol{x}_0)|\sigma^2,\phi,\boldsymbol{Z}=\boldsymbol{z}]=\left(\boldsymbol{r}+\boldsymbol{1}_n\frac{1-\boldsymbol{1}_n'\boldsymbol{R}_{\phi}^{-1}\boldsymbol{r}}{\boldsymbol{1}_n'\boldsymbol{R}_{\phi}^{-1}\boldsymbol{1}_n}\right)'\boldsymbol{R}_{\phi}^{-1}\boldsymbol{Z},$$
$$\mbox{Var}[Z(\boldsymbol{x}_0)|\sigma^2,\phi,\boldsymbol{Z}=\boldsymbol{z}]=\sigma^2\left(1-\boldsymbol{r}'\boldsymbol{R}_{\phi}^{-1}\boldsymbol{r}+\frac{{(1-\boldsymbol{1}_n'\boldsymbol{R}_{\phi}^{-1}\boldsymbol{r})}^2}{\boldsymbol{1}_n'\boldsymbol{R}_{\phi}^{-1}\boldsymbol{1}_n}\right),$$
with $\boldsymbol{r} \in \mathbb{R}^n$ the correlation vector defined as $\sigma^2\boldsymbol{r}=(\mbox{Cov}(Z(\boldsymbol{x}_0), Z(\boldsymbol{x}_j))_{1 \le j \le n}$. 

A major concern for applications of these equations is that they are conditional on the knowledge of the variance and range parameters, which is mostly unrealistic since they are estimated. This assumption yields overoptimistic prediction variances and narrower predictive intervals. This problem is made worse in the case of a small dataset where parameter estimation is sensitive to each observation. To address this issue, \citet{bachoc_estimation_2013} uses a cross-validation procedure instead of the MLE to estimate the model parameters in a more robust way, especially in the case of model misspecification. However, this approach always results in a single set of parameter values, tainted by an estimation error that is not taken into account. To remedy this, another solution is to consider the parameters as random variables, and then to quantify and finally propagate their uncertainties on the kriging model. The Bayesian approach therefore appears natural for this and leads to Bayesian kriging.

\subsection{Bayesian kriging principles}
\label{sec:2.3}

Bayesian kriging deals simultaneously with estimation and predictions by considering the parameters as random variables that must be predicted conditionally to the observed data \citep{diggle_bayesian_2002}. Bayesian kriging predictions are derived from the predictive distribution as

\begin{align*}
p_{Z(\boldsymbol{x}_0)}(Z(\boldsymbol{x}_0)|\boldsymbol{Z}=\boldsymbol{z})&=\int_{D_{\beta}{\times}D_{\sigma^2}{\times}D_{\phi}}p_{Z(\boldsymbol{x}_0),\beta,\sigma^2,\phi}(Z(\boldsymbol{x}_0),\beta,\sigma^2,\phi|\boldsymbol{Z}=\boldsymbol{z})d{\beta}d{\sigma^2}d\phi\\
& =\int_{D_{\beta}{\times}D_{\sigma^2}{\times}D_{\phi}}p_{Z(\boldsymbol{x}_0)}(Z(\boldsymbol{x}_0)|\beta,\sigma^2,\phi,\boldsymbol{Z}=\boldsymbol{z})\\ 
& \hspace{3.5cm} p_{\beta,\sigma^2,\phi}(\beta,\sigma^2,\phi|\boldsymbol{Z}=\boldsymbol{z}) d{\beta}d{\sigma^2}d\phi.
\end{align*}

The density $p_{Z(\boldsymbol{x}_0)}(Z(\boldsymbol{x}_0)|\beta,\sigma^2,\phi,\boldsymbol{Z}=\boldsymbol{z})$ is known to be a Student's $t$-density under the assumption that the prior is of the same family as the one presented at the end of this section (as demonstrated in \citet{le_interpolation_1992}), 
but the integral is usually intractable. 
In practice, it must therefore be estimated numerically by Markov chain Monte Carlo methods. 
One solution is to sample from the target distribution using a Monte Carlo approach. One such method is given in \citet{tanner_tools_1993}, and used in the geoR package \citep{ribdig01} of the  R software. A slightly different approach considers a Markov Chain for its Monte Carlo algorithm as described in \citet{gaudard_bayesian_1999} and \citet{carlin_bayesian_2009}.
So, the algorithm described by Algorithm \ref{alg:bk} is the one used in the geoR package and will be used in the following to estimate the Bayesian prediction.


\begin{algorithm}
\caption{Monte Carlo approximation for Bayesian kriging}
\label{alg:bk}
\begin{algorithmic}
\State Choose a prior specification and a position $\boldsymbol{x}_0$
\State Estimate $p_{\beta,\sigma^2,\phi}\left(\beta,\sigma^2,\phi|\boldsymbol{Z}=\boldsymbol{z}\right)$ by a MCMC method
\State $i \gets 0$
\While{$i \le M$}
\State ${\lbrace \widehat{\beta}_i, \widehat{\sigma}^2_i, \widehat{\phi}_i \rbrace} \gets \text{sample from } p_{\beta,\sigma^2,\phi}\left(\beta,\sigma^2,\phi|\boldsymbol{Z}=\boldsymbol{z}\right)$
\State $\widehat{\boldsymbol{z}}_{0,i} \gets \text{sample from } p_{Z(\boldsymbol{x}_0)}\left(Z(\boldsymbol{x}_0)\left|\widehat{\beta}_i, \widehat{\sigma}^2_i, \widehat{\phi}_i,\boldsymbol{Z}=\boldsymbol{z}\right.\right)$
\State $i \gets i+1$
\EndWhile
\State Compute the empirical mean and variance:
    \State \hskip1em $\widehat{\mathbb{E}}[Z(\boldsymbol{x}_0)|\boldsymbol{Z}=\boldsymbol{z})]  \gets \frac{1}{M}\sum_{i=1}^M \widehat{\boldsymbol{z}}_{0,i}$
    \State \hskip1em $\widehat{\mbox{Var}}[Z(\boldsymbol{x}_0)|\boldsymbol{Z}=\boldsymbol{z})]  \gets \frac{1}{M-1}\sum_{i=1}^M \left(\widehat{\boldsymbol{z}}_{0,i}-\widehat{\mathbb{E}}[Z(\boldsymbol{x}_0)|\boldsymbol{Z}=\boldsymbol{z})]\right)^2$
\State Return $\lbrace \widehat{\boldsymbol{z}}_{0,i}\rbrace_{ i \in \llbracket1,M\rrbracket},~\widehat{\mathbb{E}}[Z(\boldsymbol{x}_0)|\boldsymbol{Z}=\boldsymbol{z})], ~ \widehat{\mbox{Var}}[Z(\boldsymbol{x}_0)|\boldsymbol{Z}=\boldsymbol{z})]$
\end{algorithmic}
\end{algorithm}


$M$ is chosen so that the predictive distribution is sufficiently sampled to be approximated. For our application cases, $M=1000$.
Finally, a joint prior distribution is chosen for $\beta,\sigma^2, \phi$ that is
$$\pi(\beta,\sigma^2,\phi) \propto \frac{1}{\sigma^2}.$$
The resulting parameter space is
$$D_\beta = \mathbb{R}, D_{\sigma^2} = ]0,+\infty[, D_{\phi} = ]0,+\infty[.$$
Note that a sensitivity analysis is presented in the Appendix (Sect. \ref{ann:1}) to explain our choice of priors.

\section{Validation criteria}
\label{sec:3}

Choosing an ``optimal'' covariance model for geostatistical predictions is a classical issue in geostatistics \citep{chiles_geostatistics_2012}.

This topic has been recently studied in depth in \citet{demay_model_2022}, where different validation criteria are investigated to assess the quality of both the model predictions, the reliability of the associated prediction variances and more generally the accuracy of the whole predictive law. Depending on the number of observations available, these criteria can be computed either on a test sample separate from the training sample or, as here, by cross-validation. Their expressions, with some new adaptations, are given in this section in their leave-one-out cross-validation form. Extension to $K$-fold cross-validation or to test set cases are immediate. 

\subsection{Predictivity coefficient ($Q^2$)}
\label{sec:3.1}

The main goal of this coefficient, often called ``Nash-Sutcliffe criterion'' \citep{NashS70}, is to evaluate the predictive accuracy of the model by normalising the errors, allowing a direct interpretation in terms of explained variance. Its practical definition \citep{marioo08} is
$$ Q^2=1-\frac{\sum_{i=1}^{n}(z(\boldsymbol{x}_i) - \widehat{z}_{-i})^2}{\sum_{i=1}^{n}(z(\boldsymbol{x}_i) - \widehat{\mu})^2},$$
where $\widehat{z}_{-i}$ is the value predicted at location $\boldsymbol{x}_i$ by the model built without the $i$-th observation (the one located at $\boldsymbol{x}_i$) and $\widehat{\mu}$ is the empirical mean of the dataset. 
Its theoretical definition can be found in \citet{fekioo23}.

The $Q^2$ coefficient measures the quality of the predictions and how near they are to the observed values. Its formula is similar to the coefficient of determination used for regression (with independent observations), but estimated here in prediction (by using cross-validation residuals). The closer its value is to $1$, the better the predictions are (relatively to the observations). 

\subsection{Predictive variance adequacy ($PVA$)}
\label{sec:3.2}

This second criterion aims to quantify the quality of the prediction variances given by the kriging model. Finely studied in \citet{bac13,bachoc_estimation_2013} and \citet{demay_model_2022}, it is defined by
$$PVA = \left| \log \left(\frac{1}{n}\sum_{i=1}^{n}\frac{(z(\boldsymbol{x}_i)-\widehat{z}_{-i})^2}{\widehat{s}_{-i}^2}\right) \right|,$$
where $\widehat{s}_{-i}^2$ is the prediction variance (at location $\boldsymbol{x}_i$) of the model built without the $i$-th observation (the one located at $\boldsymbol{x}_i$). 

This coefficient estimates the average ratio between the squared observed prediction error and the prediction variance. It therefore gives an indication of how much a prediction variance is larger or smaller than the one expected. The closer the $PVA$ is to $0$, the better the prediction variances are. For example, a $PVA$ close to $0.7$ indicates prediction variances that are on average two times larger or smaller than the squared errors.

\subsection{Predictive interval adequacy ($PIA$)}
\label{sec:3.3}

The $PVA$ is a criterion of variance adequacy but does not take into account a possible skewness in the predictive distribution. In the Gaussian case (like ordinary kriging), mean and variance completely characterise the distribution. But in the case of Bayesian kriging where the predictive distribution is no longer Gaussian, the $Q^2$ and $PVA$ are not sufficient to evaluate the quality of the model and its prediction. As such, we propose a new complementary geometrical criterion called the predictive interval adequacy ($PIA$) and defined as
$$PIA=\left| \log \left( \frac{1}{n}\sum_{i=1}^{n}\frac{(z(\boldsymbol{x}_i)-\widehat{z}_{-i})^2}{\left( \widehat{q}_{0.31,-i}-\widehat{q}_{0.69,-i} \right)^2} \right) \right|,$$
where $\widehat{q}_{0.31,-i}$ (respectively $\widehat{q}_{0.69,-i}$) is the estimation of the quantile of order $0.31$ (respectively $0.69$) of the predictive distribution (at location $\boldsymbol{x}_i$) without the $i$-th observation. 

The $PIA$ has been defined to be identical to the $PVA$ for a Gaussian distribution.
However, rather than comparing squared errors to the predictive variance, it compares the width of predictive intervals with the squared errors. Another main difference is that the intervals considered by the $PIA$ are centered on the median while those of the $PVA$ are centered around the mean. Finally, an estimation of the predictive distribution is necessary to compute in practice this criterion, whereas the $PVA$ only requires the computation of predictive mean and variance.

\subsection{$\alpha$-CI plot}
\label{sec:3.4}

\begin{figure}[!ht]
\begin{center}
\includegraphics[scale = 0.4]{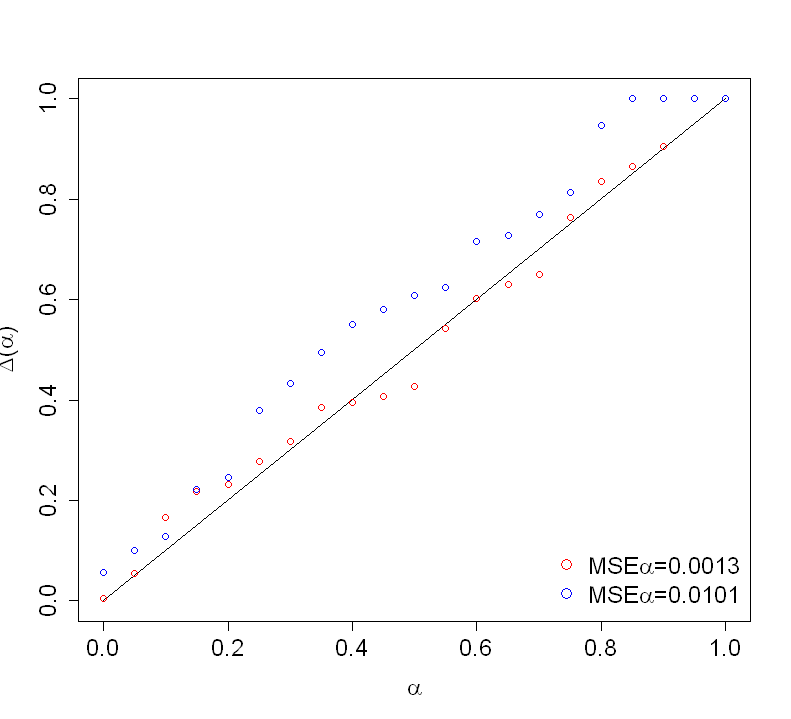}
\captionof{figure}{Example of two $\alpha$-CI plots and corresponding values of MSE$\alpha$.}
\label{fig:1} 
\end{center}
\end{figure}

The Gaussian process model allows to build predictive intervals of any level $\alpha \in ]0,1[$ written as
$$CI_{\alpha}(z(\boldsymbol{x}_i))=\left[\widehat{z}_{-i}-\widehat{s}_{-i}q_{(1+\alpha)/2}^{\mathcal{N}}\,;\,\widehat{z}_{-i}+\widehat{s}_{-i}q_{(1+\alpha)/2}^{\mathcal{N}}]\right],$$
where $q^{\mathcal{N}}_{(1+\alpha)/2}$ is the quantile of order $(1+\alpha)/2$ of the standard normal distribution. This expression is only valid if all parameters are known. For example, if the variance parameter is poorly estimated, the width of the predicted confidence intervals will not reflect what we might observe. But how can we validate a predictive interval without prior knowledge of the model parameters? The idea behind this criterion (see \citet{marioo12} and \citet{demay_model_2022}) is to evaluate empirically the number of observations falling into the predictive intervals and to compare this empirical estimation to the theoretical ones expected, with
$$\Delta_{\alpha}=\frac{1}{n}\sum_{i=1}^{n}\phi_i ~\mbox{where}~
\delta_i= \left\{
    \begin{array}{ll}
        1 & \mbox{if } z(\boldsymbol{x}_i)\in  CI_{\alpha}(z(\boldsymbol{x}_i))\\
        0 & \mbox{else.}
    \end{array}
\right.$$
This value can be computed for varying $\alpha$, and can then be visualised against the theoretical values, yielding what \citet{demay_model_2022} calls the $\alpha$-CI plot, with an example given in Fig. \ref{fig:1}.

Similarly to the $PIA$, the $\alpha$-CI plot must be adapted to the Bayesian kriging since the posterior distribution is not Gaussian. We therefore introduce a slightly different criterion based on the quantiles of the predictive distribution. 
More precisely, this modified $\alpha$-CI plot relies now on credible intervals defined as
$$\widetilde{CI}_{\alpha}(z(\boldsymbol{x}_i))=\left[\widehat{q}_{\frac{1-\alpha}{2}};\widehat{q}_{\frac{1+\alpha}{2}}\right],$$
where $\widehat{q}_{\frac{1-\alpha}{2}}$ (respectively $\widehat{q}_{\frac{1+\alpha}{2}}$) is the estimation of the quantile of order $\frac{1-\alpha}{2}$ (respectively $\frac{1+\alpha}{2}$) of the predictive distribution (at location $\boldsymbol{x}_i$) of the model built without the $i$-th observation.
Once again, we obtain a criterion that is identical for both methods when the predictive distribution is Gaussian.

\subsection{Mean Squared Error $\alpha$ ($MSE\alpha$)}
\label{sec:3.5}

Finally, to summarise the $\alpha$-CI plot, we also introduce a quantitative criterion called ``Mean Squared Error $\alpha$'' and defined as
$$MSE\alpha = \frac{1}{n_{\alpha}}\sum_{j=1}^{n_{\alpha}}(\Delta_{\alpha_j}-\alpha_j)^2,$$
where the considered levels $\alpha$ are discretized over $]0,1[$ in $n_\alpha$ possible values.
In practice a regular discretization will be considered to compute $MSE\alpha$. The closer this criterion is to $0$, the better the predictive/credible intervals are on average. To illustrate the values taken by the criterion, Fig. \ref{fig:1} gives the $\alpha$-CI plot corresponding to a ``good'' and ``bad'' model fitting. In this graph, the bad model yields a $MSE\alpha$ of $0.0101$ against $0.0013$ for a model with more accurate predictive intervals.
In the context of dismantling and decommissioning of nuclear sites, a $MSE\alpha$ of $0.01$ will be considered to correspond to a model with wrong predictive intervals, while a model with a $MSE\alpha$ of $0.001$ will be deemed to have correct predictive intervals. Similarly to the $PVA$, the $MSE\alpha$ does not explain if the poorly fitted predictive intervals are due to badly centered predictive intervals or if the predictive variance was badly estimated (and whether or not this variance was underestimated or overestimated). This criterion must therefore be used in conjunction with the previous criteria to better assert the model qualities and weaknesses. Finally, this criterion also offers a quantitative tool for comparing different models if the $\alpha$-CI plots do not allow to clearly distinguish the performances of competing models.
This will be illustrated in particular in the numerical tests in Sect. \ref{sec:4.2} (Fig. \ref{fig:8}).

The different aforementioned criteria provide complementary information to evaluate the prediction quality of the kriging model, either in terms of mean, variance or predictive/credible intervals. They will be used in the following to compare the performance of ordinary and Bayesian kriging.

\section{Numerical tests and results}
\label{sec:4}

Our goal is to compare Bayesian and ordinary kriging (the latter being the more commonly used kriging method)\footnote{The R code corresponding to these tests is given in \url{https://gitlab.com/biooss/r-code-for-wieskotten-et-al-2023-paper}}. To do so, the different criteria mentioned in Sect. \ref{sec:3} will be computed on datasets (i.e., samples of observations), coming from different models, of different sizes. Parameter estimation results are not discussed further here, but an analysis is given in Appendix \ref{ann:2}.

\subsection{Datasets from 2D Gaussian process simulations}
\label{sec:4.1}

First, we consider samples simulated from an analytical Gaussian process model with known parameters. More precisely, the samples are simulated in the input space $[0,10]^2$ from a Gaussian process with an exponential covariance (i.e., the Matérn covariance of Eq. \eqref{eq:matern} with $\nu=0.5$) and the parameters
$$\beta = 0.5,\sigma^2=0.1,\phi=4.5 .$$
We simulate datasets of different sizes, varying from $16$ to $81$ observations, sampled on a square grid in the input space. Here, the sampling designs will be regular squared grids.
This choice is made to comply with the application purpose which deals with D\&D constraints of buildings.
Indeed, most of the times, the radiological measurements inside buildings are made regularly (equidistant location) along lines of investigations (see, e.g. \citet{cet14} and \citet{epri}). For each size, the process is repeated $100$ times with independent random Gaussian process simulations.

\begin{figure}[!ht]
\begin{center}
\includegraphics[width=1.1\textwidth]{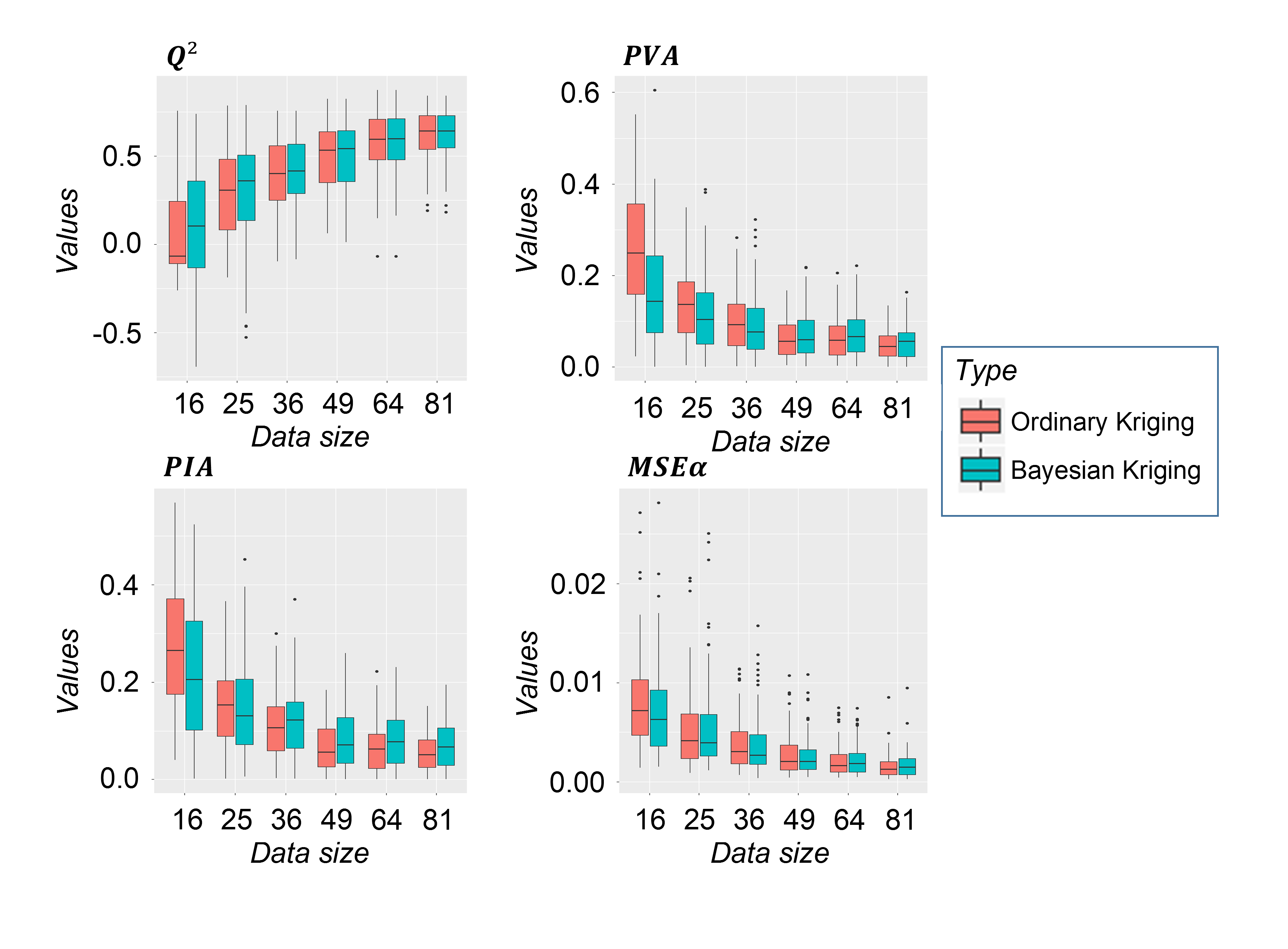}
\captionof{figure}{Distribution of validation criteria ($Q^2$, $PVA$, $PIA$, and $MSE\alpha$) w.r.t. the size of datasets, for Gaussian process simulation datasets.}
\label{fig:2} 
\end{center}
\end{figure}

For each dataset, Bayesian and ordinary kriging models are estimated and the different validation criteria are computed by cross-validation. Every kriging predictions (Bayesian and ordinary) are made with the  R package  geoR \citep{ribdig01}. Results are given in Fig. \ref{fig:2} with boxplots (corresponding to the $100$ random replicates) w.r.t. the dataset sizes.

The results for the validation criteria indicate that Bayesian kriging performs better in terms of both mean and prediction variance for small sample sizes. More precisely, Bayesian kriging outperforms ordinary kriging on most criteria for datasets with less than $40$ observations (with the exception of the $PIA$, where for $36$ observations, ordinary kriging outperforms Bayesian kriging). 

More precisely, if we first look at the median values of $Q^2$ estimation, these increase from $-0.07$ to $0.64$, according to the data size, for ordinary kriging.
Bayesian kriging gives better $Q^2$ for smaller datasets, starting from a median value of $0.10$ up to $0.64$. 
For a fixed sample size, the dispersion of $Q^2$ is quite similar between both kriging methods (for example, we have a standard deviation of $0.21$ for both methods for $36$ observations).

Regarding the median of $PVA$, the value range from $0.25$ to $0.04$ for ordinary kriging, compared to $0.14$ to $0.06$ for Bayesian kriging. For the $PIA$, the results are identical for ordinary kriging, but Bayesian kriging performs slightly worse, starting at $0.21$ up to $0.05$. We can also see that the dispersion of $PIA$ and $PVA$ estimates is different for small datasets between both kriging methods. 
This is explained by the fact that $PVA$ and $PIA$ are sensitive to the parameter estimation process. Since the number of observations is low, maximum likelihood estimations are not robust, yielding large variations in parameter estimations, and therefore in $PVA$ and $PIA$ estimations. Finally, we observe that for datasets larger or equal to $49$, Bayesian kriging seems to perform slightly worse than ordinary kriging. 

The $MSE\alpha$ graph shares similarities with the other graphs, since predictive and credible intervals both depends on prediction mean and variance. For the ordinary kriging, the median $MSE\alpha$ goes from $0.0072$ to $0.0012$, while for Bayesian kriging the values are lower, from $0.0063$ to $0.0015$. The evolution observed is similar between the $PVA$ and $PIA$, with Bayesian kriging yielding better results for smaller datasets.

It can also be noted that for larger datasets, Bayesian kriging yields slightly worse results. It can therefore be argued that Bayesian kriging becomes less advantageous and relevant for datasets with more than $40$ observations. Note that $Q^2$ values are also extremely low for 49 observations or fewer, but again this is to be expected for very small datasets.

\subsection{Datasets from a 2D deterministic function}
\label{sec:4.2}

\begin{figure}[!ht]
\begin{center}
\includegraphics[width=0.7\textwidth]{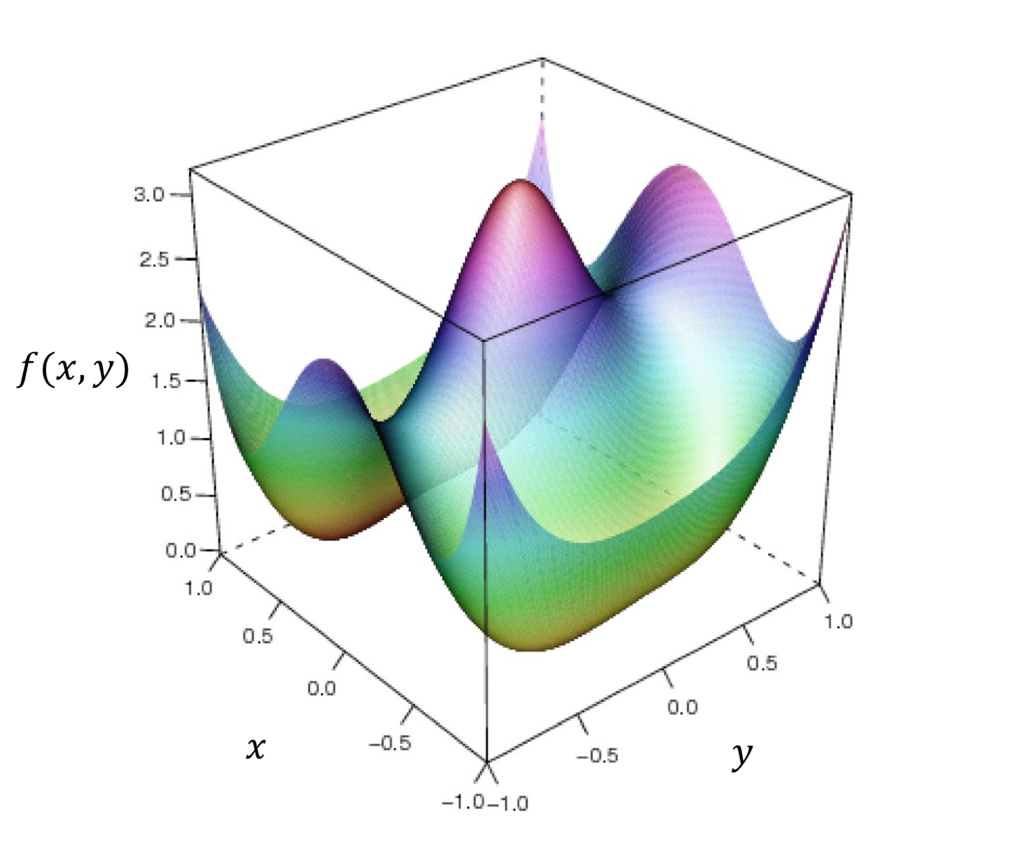}
\caption{Illustration of the deterministic function $f$ \citep{iooss_numerical_2010}.}
\label{fig:3} 
\end{center}
\end{figure}

In order to test the kriging models on cases that do not fall within the theoretical framework of the Gaussian process hypothesis, we consider a sample coming from the following two-dimensional deterministic function \citep{iooss_numerical_2010}
\begin{equation}\label{eq:detfct}
f(x,y)=\frac{e^x}{5}-\frac{y}{5}+\frac{y^6}{3}+4y^4-4y^2+\frac{7x^2}{10}+x^4+\frac{3}{4x^2+4y^2+1},
\end{equation}
where $(x,y)$ are the function inputs. Figure \ref{fig:3} shows this function over the $D=[-1,1]^2$ input space.

We consider two steps for studying this test function. 
First, the validation criteria are used to compare the results obtained by using different covariance functions in order to identify the most appropriate one for the dataset (as done in \citet{demay_model_2022}).

\begin{figure}
\begin{subfigure}[t]{0.5\textwidth}
  \centering
  \includegraphics[width=\textwidth]{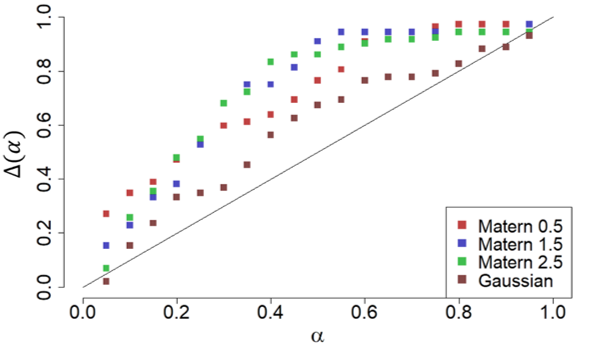}
  \caption{Ordinary kriging.}
  \label{fig:4a}
\end{subfigure}%
\begin{subfigure}[t]{0.5\textwidth}
  \centering
  \includegraphics[width=\textwidth]{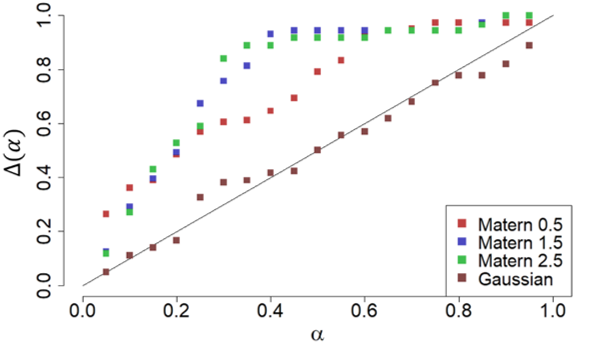}
  \caption{Bayesian kriging.}
  \label{fig:4b}
\end{subfigure}
\caption{$\alpha$-CI plots for the ordinary and Bayesian kriging with different covariances functions, on the sample of $n=144$ observations of function $f$.}
\label{fig:4}
\end{figure}

Then, a regular squared grid is considered to sample the input space, composed of $144$ observations.
On this dataset, the ordinary kriging model is fitted with different covariance functions, namely three Mat\'ern covariances and the Gaussian covariance with a nugget effect for the latter of $10^{-6}$ (to improve the numerical stability of the covariance matrix inversion). For each of these covariances, the validation criteria are estimated by a cross validation process.
The results are presented in Tab. \ref{tab:1} for ordinary kriging, in Tab. \ref{tab:2} for Bayesian kriging and in Fig. \ref{fig:4}.

The main goal of this procedure is to better identify the covariance, so that this choice has no concern for the rest of our study. Therefore, a dataset of $144$ observations is used to ensure a good analysis of the covariance function through the use of the aforementioned validation criteria.

\begin{table}[!ht]
\centering
\begin{tabular}{ | l | c | c | c | c |}
    \hline
   {\bf Covariance} & $Q^2$ & $PVA$ & $PIA$ & $MSE\alpha$ \\ \hline
   Matérn $1/2$ & $0.95$ & $0.99$ & $0.98$ & $0.056$ \\ \hline
   Matérn $3/2$ & $0.99$ & $0.91$ & $0.90$ & $0.073$ \\ \hline
   Matérn $5/2$ & $1.00$ & $0.65$ & $0.63$ & $0.073$ \\ \hline
   Gaussian & $1.00$ & $0.05$ & $0.07$ & $0.011$ \\ \hline
   \end{tabular}
    \caption{Validation criteria for the ordinary kriging with different covariance functions, on the sample of $n=144$ observations of function $f$.}
    \label{tab:1}
\end{table}

\begin{table}[!ht]
\centering
\begin{tabular}{ | l | c | c | c | c |}
    \hline
   {\bf Covariance} & $Q^2$ & $PVA$ & $PIA$ & $MSE\alpha$ \\ \hline
   Matérn $1/2$ & $0.95$ & $1.09$ & $1.06$ & $0.061$ \\ \hline
   Matérn $3/2$ & $0.99$ & $1.62$ & $1.60$ & $0.106$ \\ \hline
   Matérn $5/2$ & $1.00$ & $1.58$ & $1.55$ & $0.106$ \\ \hline
   Gaussian & $1.00$ & $0.13$ & $0.16$ & $0.002$ \\ \hline
   \end{tabular}
    \caption{Validation criteria for the Bayesian kriging with different covariance functions, on the sample of $n=144$ observations of function $f$.}
    \label{tab:2}
\end{table}

The results show that, in this case, a Gaussian covariance function is the most appropriate covariance function w.r.t. to the different criteria. This result is not surprising since the test function is smooth and shows large correlations between observations. 
Although the differences between $Q^2$ are very small between the Gaussian and Matérn models (except for the Matérn $1/2$ model), significant differences appear for the $PVA$ and $PIA$.
These differences become smaller for the $MSE\alpha$. This shows the importance of using simultaneously various criteria for a better assessment of the model performance and accuracy.

Once our covariance model is chosen (the Gaussian one in this case), we can apply a similar test protocol than in Sect. \ref{sec:4.1}. In order to generate datasets, we have to slightly modify the protocol. Since the function is deterministic, choosing a specific geometry for a fixed dataset size will not allow to generate different datasets. Therefore we discard here the regular grid and choose to sample random positions in the input space. It allows us to generate different datasets while considering the same deterministic function, even though such random sampling would not be recommended in practice. The observed dispersion in the results of this section is affected by that choice. This sampling is repeated $100$ times for each dataset size, up to $150$ observations.

\begin{figure}[!ht]
\begin{center}
\includegraphics[width=1.1\textwidth]{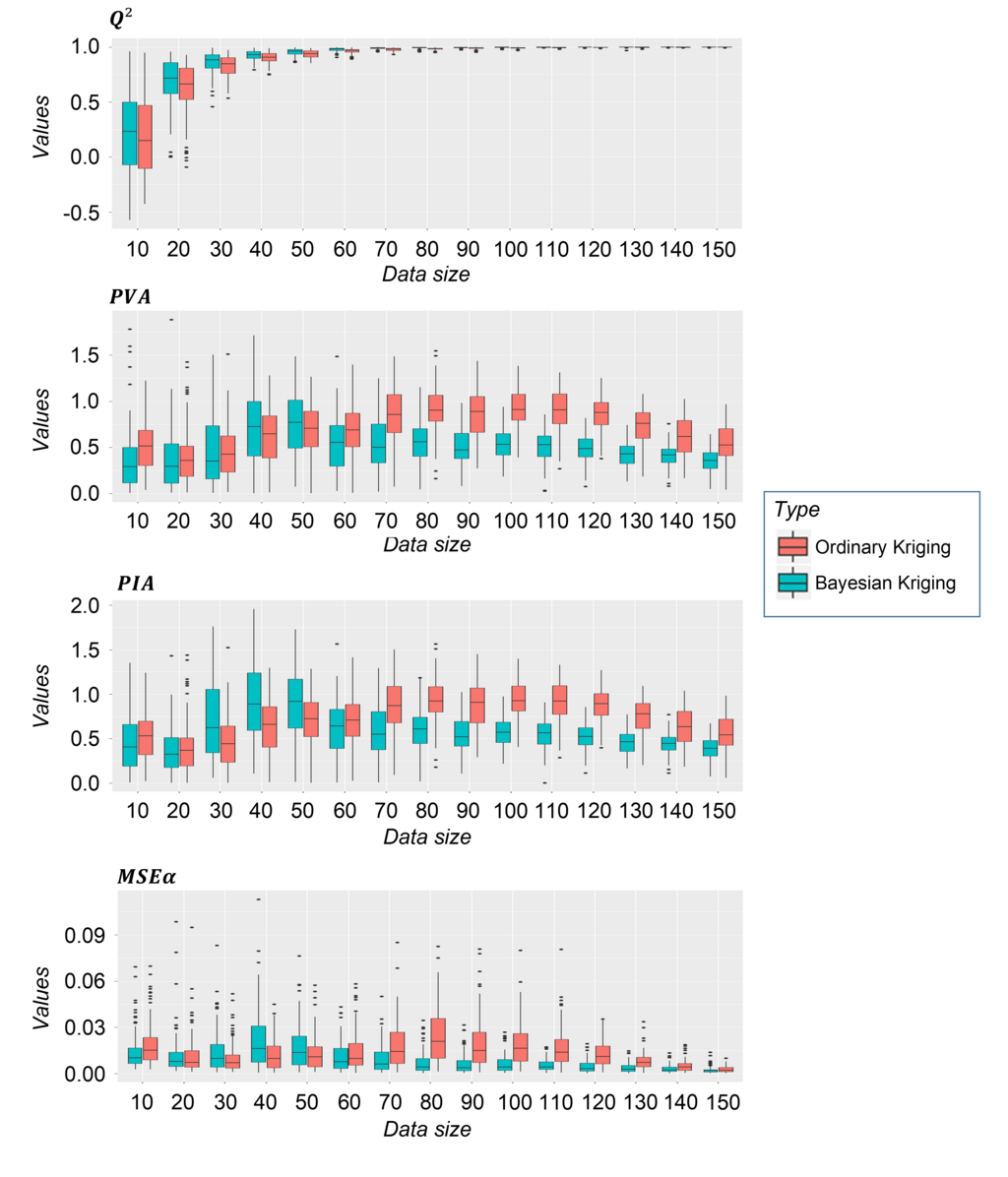}
\captionof{figure}{Distribution of validation criteria ($Q^2$, $PVA$, $PIA$, and $MSE\alpha$) w.r.t. the size of datasets, for the deterministic function $f$.}
\label{fig:5}
\end{center}
\end{figure}

The results are presented in Fig. \ref{fig:5}.
The values of the $Q^2$ criterion lead to the same conclusions as for the data from Gaussian process trajectories, in the previous section. We again find better performance with Bayesian kriging, especially for small sample sizes. Note that we have higher $Q^2$'s than for the previous test case due to the high regularity of the function $f$.


Significant differences arise with the $PVA$, $PIA$ and $MSE\alpha$ criteria. 
Indeed, these criteria do not decrease steadily and monotonically with the number of observations. Moreover, they behave differently depending on the type of kriging. More precisely, for Bayesian kriging, the $PVA$, $PIA$ and $MSE\alpha$ increase between $20$ observations and $50$ observations, before decreasing, whereas they keep increasing for ordinary kriging.
For datasets made of $50$ observations or less, Bayesian kriging seems to under-perform when compared to ordinary kriging but outperform ordinary kriging for more than $50$ observations. Still, once the size of the datasets exceed $80$ observations, we observe similar results to those obtained with the simulated datasets.


To explain these results, we recall that the initial assumption whereby the function $f$ is a trajectory of a Gaussian process is not verified here, at least for datasets of $50$ or less observations.  
It is therefore possible to obtain poorer criteria as the dataset size increases. 
We still get good prediction accuracy, since the median of the $Q^2$ criterion stays between $0.7$ and $1$ for all dataset sizes and kriging methods, but the predicted variances do not seem to be very accurate, yielding poorly estimated predictive and credible intervals. We can observe that once the dataset size exceeds $80$ observations, the evolution of the validation criteria shows that the initial assumption is now valid.


In conclusion, Bayesian kriging outperforms on average ordinary kriging in this case where the initial assumption of a Gaussian random field is not true.
Caution is still advised, since in some cases ordinary kriging seems to perform better than Bayesian kriging, as illustrated with the $n=40$ or $n=50$ observations' dataset. The conclusion obtained in Sect. \ref{sec:4.1} cannot be made identically here, because for small data sets, Bayesian kriging does not seem to consistently give better validation criteria.

\section{Real application case: G3's dataset}
\label{sec:5}

\begin{figure}[!ht]
\begin{center}
\includegraphics[scale = 0.4]{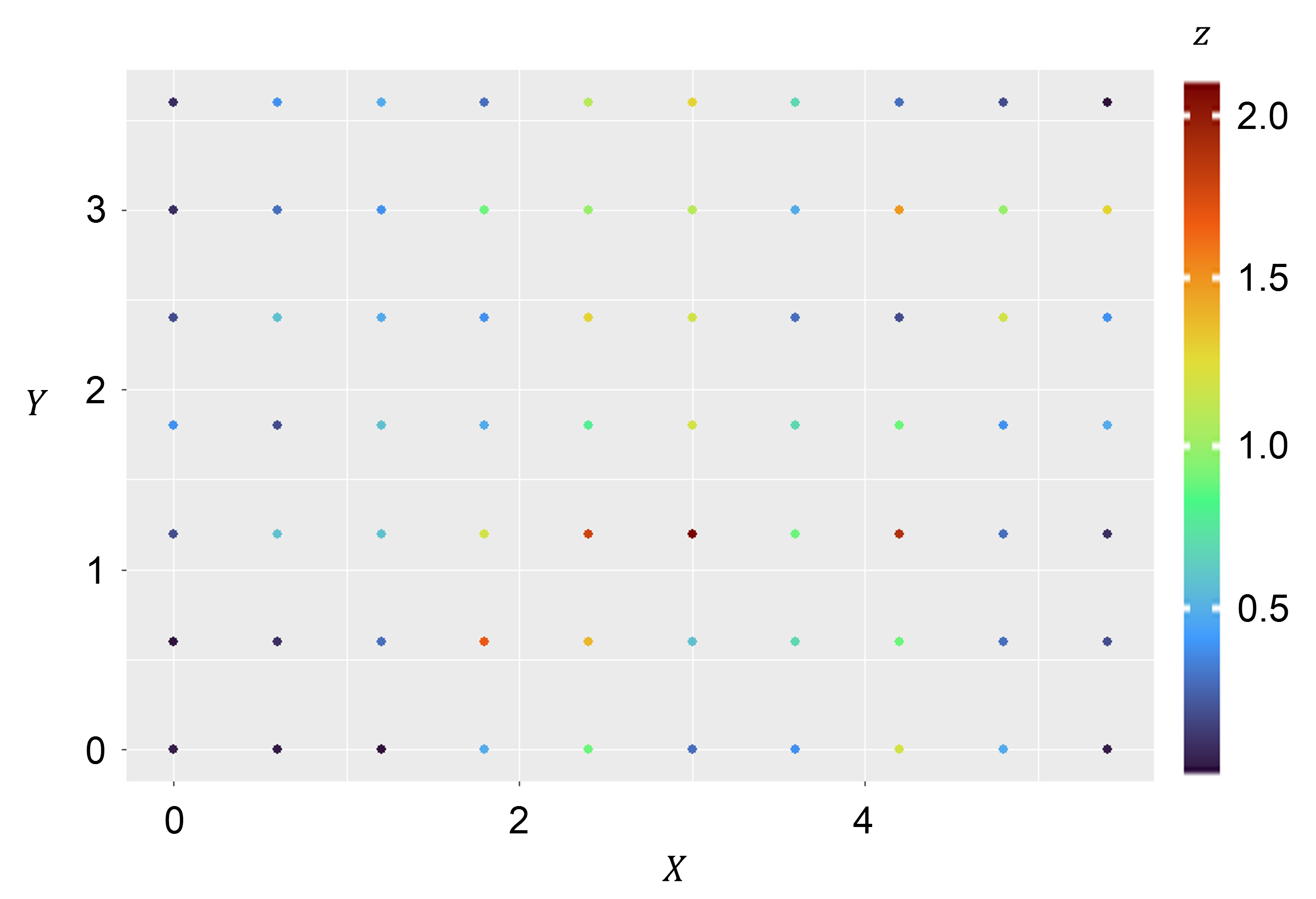}
\captionof{figure}{Mapping of G3 observations.}
\label{fig:6}
\end{center}
\end{figure}

This dataset is made of $70$ observations of radioactivity measurements coming decommissioning project of the CEA Marcoule G3 reactor \citep{G3reactor}.
They are sampled in the input domain $[0,6] \times [0,4]$. The dataset is mapped in Fig. \ref{fig:6}.

\begin{figure}[!ht]
\begin{center}
\includegraphics[scale = 0.1]{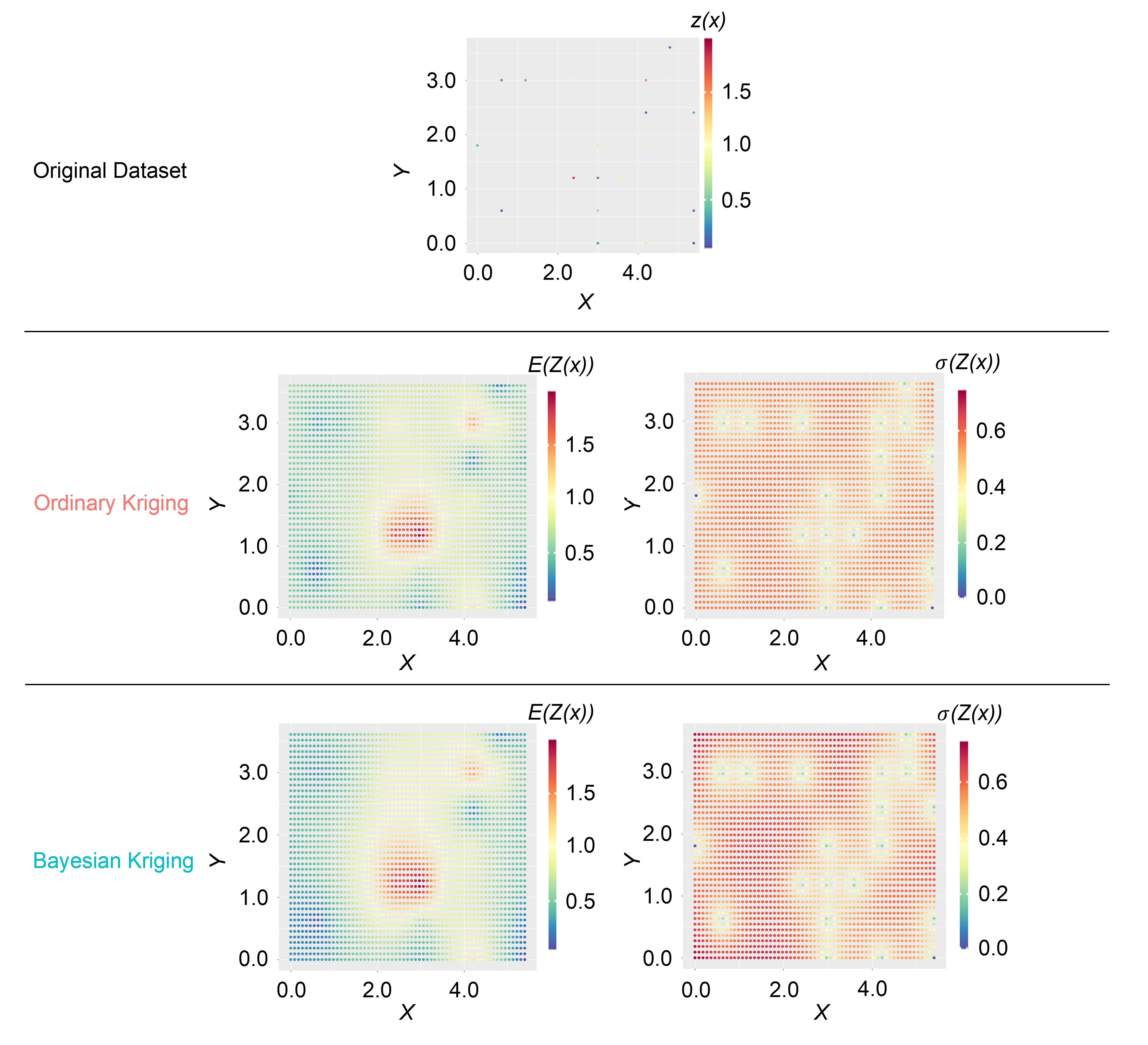}
\captionof{figure}{Predictions for $20$ observations sampled from the original dataset with ordinary and Bayesian kriging.}
\label{fig:7}
\end{center}
\end{figure}

Figure \ref{fig:7} shows the predictions of Bayesian kriging and ordinary kriging for a given dataset of $n=20$ observations (randomly sampled from the original data set).
More precisely, the prediction maps obtained with ordinary and Bayesian kriging with an exponential covariance for both models are given. The figure also highlights the differences between both predictions. A small difference between predicted standard deviation appears, since they are much higher for Bayesian kriging. This is explained by the fact that for a small number of observations, Bayesian kriging takes more uncertainty into account, resulting in higher prediction variances. In the practice of D\&D projects, this can have a direct impact since the estimates (or more precisely the upper quantiles or margins given by the predictive law) will be more conservative. Note that as we increase the sample size, the differences between the Bayesian and ordinary kriging maps are no longer visible.
Indeed, the uncertainty of parameter estimation (only taken into account by the Bayesian kriging) becomes negligible in front of the interpolation uncertainty (common to the two kriging methods).

\begin{figure}
\begin{subfigure}[t]{0.5\textwidth}
  \centering
  \includegraphics[width=1\textwidth]{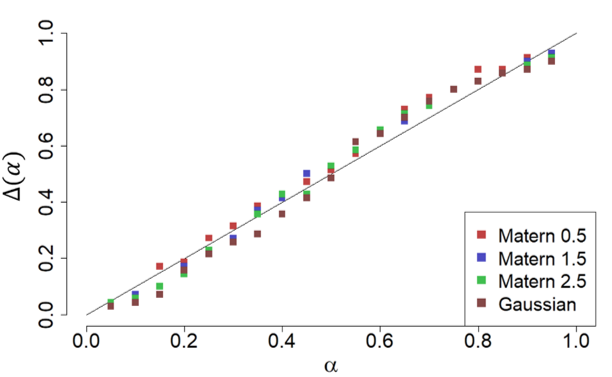}
  \caption{Ordinary kriging.}
  \label{fig:8a}
\end{subfigure}%
\begin{subfigure}[t]{0.5\textwidth}
  \centering
  \includegraphics[width=1\textwidth]{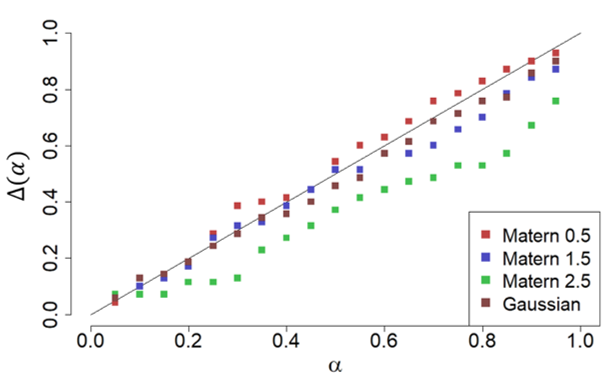}
  \caption{Bayesian kriging.}
  \label{fig:8b}
\end{subfigure}
\caption{$\alpha$-CI plots for the ordinary and Bayesian kriging with different covariance functions, on the G3 sample of $n=70$ observations.}
\label{fig:8}
\end{figure}

Let us now examine the effects of varying sample sizes and covariance models. A similar test protocol as in Sect. \ref{sec:4} is applied to assess the behaviour of kriging models according to $n$. 
First, let us consider ordinary kriging for different covariance functions, applied to the initial set of $70$ observations. The validation criteria estimated by cross-validation are given in Tab. \ref{tab:3} and Fig. \ref{fig:8}. For Bayesian kriging, they are given in Tab. \ref{tab:4} and Fig. \ref{fig:8}.
The results indicate that the  Matérn $1/2$ model is the best choice in regards of our different criteria since it maximizes the $Q^2$ criterion while minimizing both $PVA$ and $PIA$ criteria (it also performs well for the $MSE\alpha$ criterion, while not being the function minimizing it overall). Therefore only the Matérn $1/2$ covariance function is considered.

\begin{table}[!ht]
\centering
\begin{tabular}{ | l | c | c | c | c |}
    \hline
   {\bf Covariance} & $Q^2$ & $PVA$ & $PIA$ & $MSE\alpha$ \\ \hline
   Matérn $1/2$ (exponential) & $0.37$ & $0.06$ & $0.07$ & $0.0015$ \\ \hline
   Matérn $3/2$ & $0.33$ & $0.12$ & $0.14$ & $0.0010$ \\ \hline
   Matérn $5/2$ & $0.31$ & $0.14$ & $0.15$ & $0.0014$ \\ \hline
   Gaussian & $0.24$ & $0.16$ & $0.18$ & $0.0021$ \\ \hline
\end{tabular}
\caption{Validation criteria for the ordinary kriging with different covariance functions, on the G3 sample of $n=70$ observations.}
\label{tab:3}
\end{table}

\begin{table}[!ht]
\centering
\begin{tabular}{ | l | c | c | c | c |}
    \hline
   {\bf Covariance} & $Q^2$ & $PVA$ & $PIA$ & $MSE\alpha$ \\ \hline
   Matérn $1/2$ (exponential) & $0.38$ & $0.12$ & $0.07$ & $0.0013$ \\ \hline
   Matérn $3/2$ & $0.20$ & $0.51$ & $0.55$ & $0.0028$ \\ \hline
   Matérn $5/2$ & $0.16$ & $1.19$ & $1.25$ & $0.0284$ \\ \hline
   Gaussian & $0.15$ & $0.36$ & $0.40$ & $0.0015$ \\ \hline
\end{tabular}
\caption{Validation criteria for the Bayesian kriging with different covariance functions, on the G3 sample of $n=70$ observations.}
\label{tab:4}
\end{table}

To generate multiple datasets, we resample without replacement datasets of various sizes $n=20,30,40,50,60,70$, with the last one being the original dataset. Once again, the process is repeated $100$ times for each sample size (except for $70$ observations) and for each sample a cross-validation is applied to estimate the validation criteria.

\begin{figure}[!ht]
\begin{center}
\includegraphics[width=1.1\textwidth]{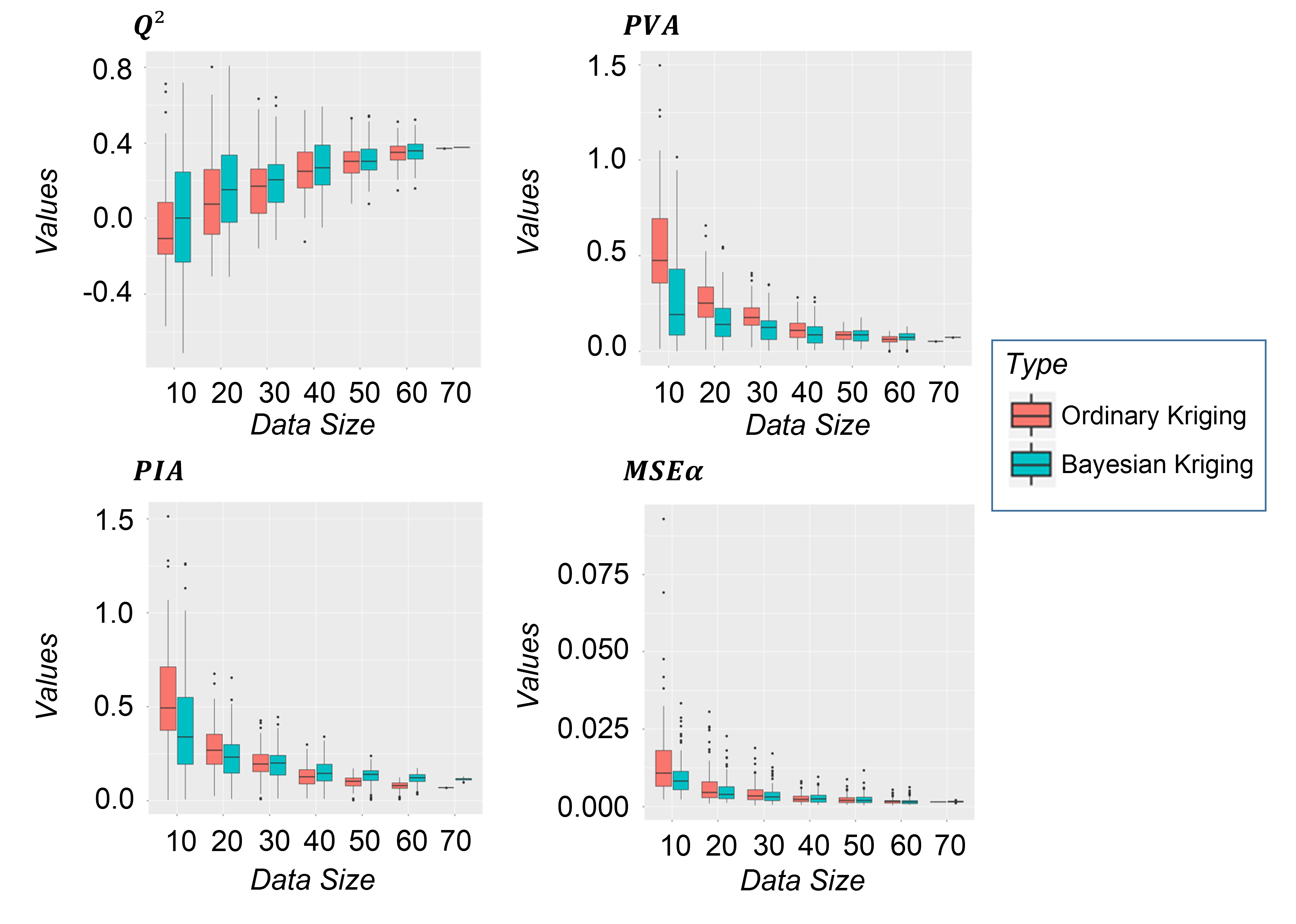}
\captionof{figure}{Distribution of validation criteria ($Q^2$, $PVA$, $PIA$, and $MSE\alpha$) w.r.t. the size of datasets, for the G3 dataset.}
\label{fig:9}
\end{center}
\end{figure}

The obtained results are summarised in Fig. \ref{fig:9}. For the $Q^2$ criterion, the median values increase from about $0$ ($n=10$) to $0.38$ ($n=70$) for both kriging methods.
Slightly higher results are obtained for Bayesian kriging, especially for small sample sizes.
The dispersion of $Q^2$ is similar between the two kriging methods. 
The obtained $Q^2$ estimates here are very low, which normally means that the model is not predictive enough.
As our objective is only to compare the kriging methods, this problem is not further investigated here.

Regarding the $PVA$, the median values decrease from $0.47$ to $0.16$ for ordinary kriging, compared to much lower values for Bayesian kriging, namely from $0.19$ to $0.06$
For the $PIA$, the values are very close to the ones of $PVA$.
For the $MSE\alpha$, the median values go from $0.011$ to $0.0017$ for the ordinary kriging, against $0.008$ down to $0.0017$ for Bayesian kriging. 
Once again, Bayesian kriging yields better results, especially for smaller datasets.
The results of both methods then become almost identical for datasets of $40$ or more observations.
This is especially visible for the $MSE\alpha$.

We can also remark that the variance of each validation criterion is reduced as the datasets size grows. This is both explained by the larger datasets, but also by our protocol, since observations are randomly drawn without replacement among the original $70$ observations.
As a result, the samples differ less and less as the dataset sizes increases. 

\section{Discussion and Conclusions}
\label{sec:6}

In conclusion, the use of Bayesian kriging for spatial interpolation of datasets in support of decommissioning and dismantling projects shows promising results. 
Its main advantage is that it allows to take into account the uncertainty of the parameters of the kriging model. 
The results given in the three application cases show that on average, Bayesian kriging outperforms ordinary kriging. Still, the second case (dealing with a deterministic function) gives a clear and interesting counter-example. Even though this result could be explained with the fact that the Gaussian assumption is not verified, it advocates for cautious use of Bayesian kriging. As the sample size increases, ordinary kriging, less computationally expensive, is then preferable for large datasets. 
Bayesian kriging has also the drawback of requiring a prior specification, which is often difficult to choose and can strongly influence the predictions. Therefore, the use of Bayesian kriging should be restricted to smaller datasets or cases in which prior information on parameters is well known.

Another important advantage of Bayesian kriging is that it allows to evaluate the information brought by the data on the parameter characterization (e.g., by comparing their prior and posterior distributions), and can share the prediction uncertainty between the data interpolation uncertainties and the one coming from the parameters' uncertainties.
It then allows to judge if the latter uncertainty is negligible compared to the former in order to bring some confidence in this statistical tool to the user.
Another fruitful perspective is that the evolution of the posterior distribution could be used for defining a new design of experiments, allowing to compare the information brought by new observations.

In our work we did not use the nugget effect as a modelling tool but only as a regularisation of the Gaussian covariance function. Future works will aim at adding this parameter to the model. This could be taken further by considering a heteroscedastic model \citep{ng_bayesian_2012}, since the usual nugget effect is formulated as a homoscedastic model. This could be extremely useful and show promising results in the framework of D\&D of nuclear sites since radioactive measurements are prone to varying measurement uncertainties, depending on the measuring technique.

The results presented in this paper also show that the main differences between the two kriging methods are in the prediction variances, which are often larger with Bayesian kriging. This can lead to predictions with more conservative associated uncertainties, potentially increasing the difficulty of decision making. However, this disadvantage must be put into perspective in the framework of D\&D projects, because in this context it is preferable, for safety reasons, to overestimate contamination rather than underestimate it.


\appendix
\section{Sensitivity analysis to the prior distribution of parameters}
\label{ann:1}

The choice of prior specifications is a complicated step in Bayesian analysis. We therefore conduct a sensitivity analysis to justify our use of an improper prior on the mean and variance parameters. 
Note that the range will not be described here, since no usual specification is available. 

First, it could be argued that the prior on the parameter $\beta$ is chosen improper since this choice is implicitly made in ordinary kriging:
$$\pi(\beta) \propto 1.$$
Second, for the variance parameter $\sigma^2$, several choices for priors can be considered. To give a quick overview of our test protocol, we used a simulated dataset, defined as random trajectories of the same Gaussian process model as in Sect. \ref{sec:4.1}. 
An initial dataset of $16641$ observations is simulated, on which the parameters $\beta_\text{init}$ and $\sigma^2_\text{init}$ are estimated. 
These estimations will be considered as reference values for our prior specifications. From these $16641$ observations, samples of $n=20$ and $n=50$ observations are randomly drawn.
This sampling is then repeated $100$ times, generating a total of $200$ datasets. 
Then, for each dataset, the Bayesian kriging is applied considering five different prior specifications:
\begin{enumerate}
    \item vague with $$\pi(\beta,\sigma^2) \propto \frac{1}{\sigma^2},$$
    \item correctly centred and informative with $$\sigma^2 \sim \text{ Scaled-Inv-}\chi^2(\sigma^2_\text{init},n) \text{ and } \beta|\sigma^2 \sim \mathcal{N}(\beta_\text{init}, \frac{\sigma^2}{n}),$$
    \item incorrectly centred and informative with $$\sigma^2 \sim \text{ Scaled-Inv-}\chi^2(3\sigma^2_\text{init},n) \text{ and } \beta|\sigma^2 \sim \mathcal{N}(3\beta_\text{init}, \frac{\sigma^2}{n}),$$
    \item correctly centred and non-informative with $$\sigma^2 \sim \text{ Scaled-Inv-}\chi^2(\sigma^2_\text{init},\frac{n}{3}) \text{ and } \beta|\sigma^2 \sim \mathcal{N}(\beta_\text{init}, \frac{\sigma^2}{n}),$$
    \item incorrectly centred and non-informative with $$\sigma^2 \sim \text{ Scaled-Inv-}\chi^2(3\sigma^2_\text{init},\frac{n}{3}) \text{ and } \beta|\sigma^2 \sim \mathcal{N}(3\beta_\text{init}, \frac{\sigma^2}{n}).$$
\end{enumerate}

\begin{figure}[!ht]
\begin{center}
\includegraphics[width=1.1\textwidth]{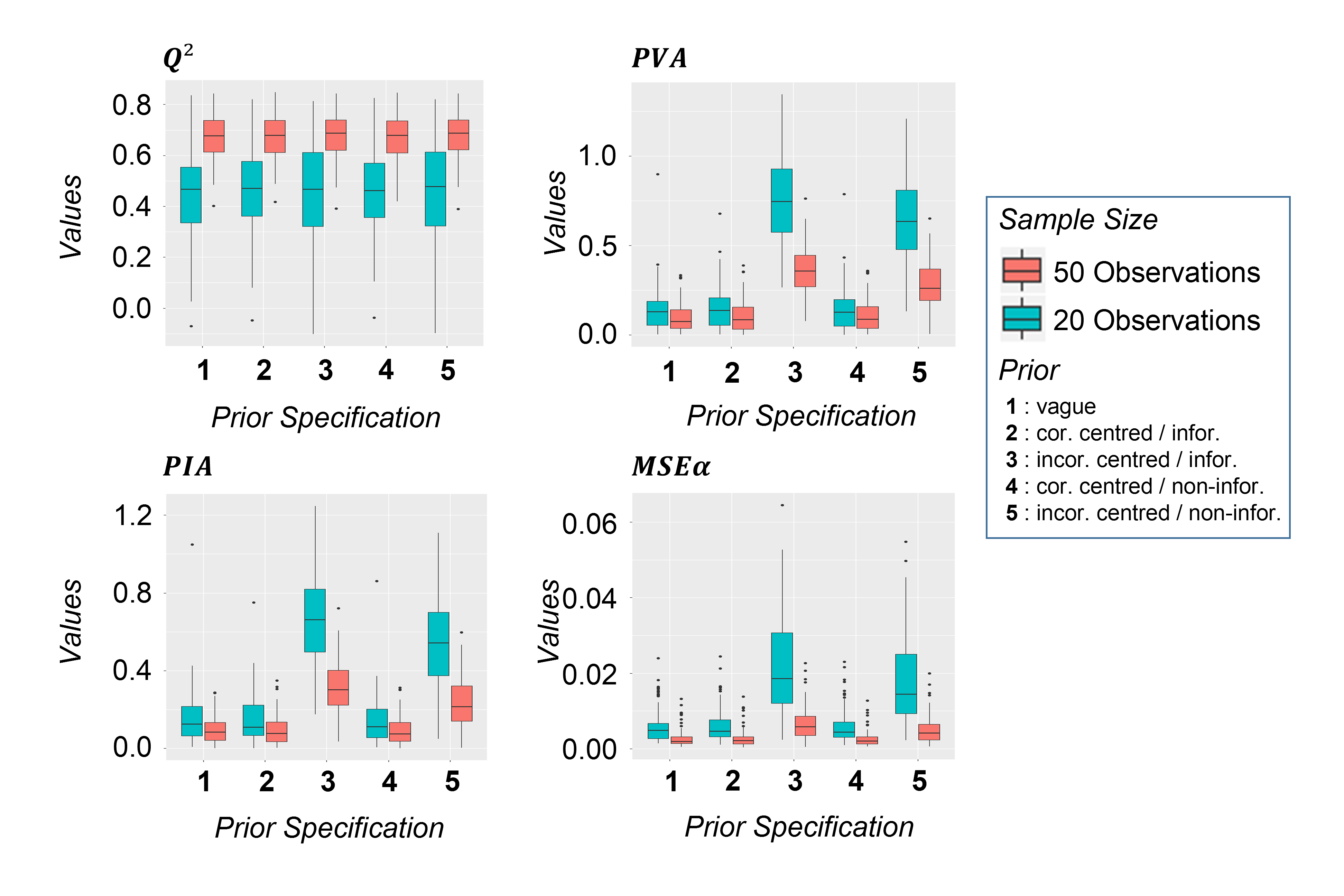}
\captionof{figure}{Distribution of validation criteria ($Q^2$, $PVA$, $PIA$, and $MSE\alpha$) w.r.t. to the prior specification.}
\label{fig:10}
\end{center}
\end{figure}

For each prior specification, Bayesian kriging is combined with cross-validation to estimate validation criteria.
The obtained results are given in Fig. \ref{fig:10}.
First, we observe that the $Q^2$ criterion is not sensitive to the prior specification. This is expected since the prediction performances depend mostly on the number of observations and on the geometry of the dataset. On contrary, the $PVA$ and $PIA$ criterion are very sensitive to the prior specification since prediction variance highly depends on parameter estimation. A vague prior allows to mitigate the bias introduced with an incorrectly centred prior, as case 3 shows a worse result than case 5. We can also see that even with a correctly centred and informative prior (case 2), the gains in parameter estimation are small if we compare it to a vague specification (case 1). 

In conclusion, the choice of a vague or improper prior is reasonable, as the improvements provided by a correctly specified prior do not seem good enough in comparison to the pitfall of a bad prior specification. These results are also similar to the one obtained by \citet{helbert_assessment_2009}.

\section{Complementary results on covariance parameter estimates}
\label{ann:2}

\subsection{Parameter estimation on simulated datasets with increasing sizes}

\begin{figure}[!ht]
\begin{center}
\includegraphics[width=1\textwidth]{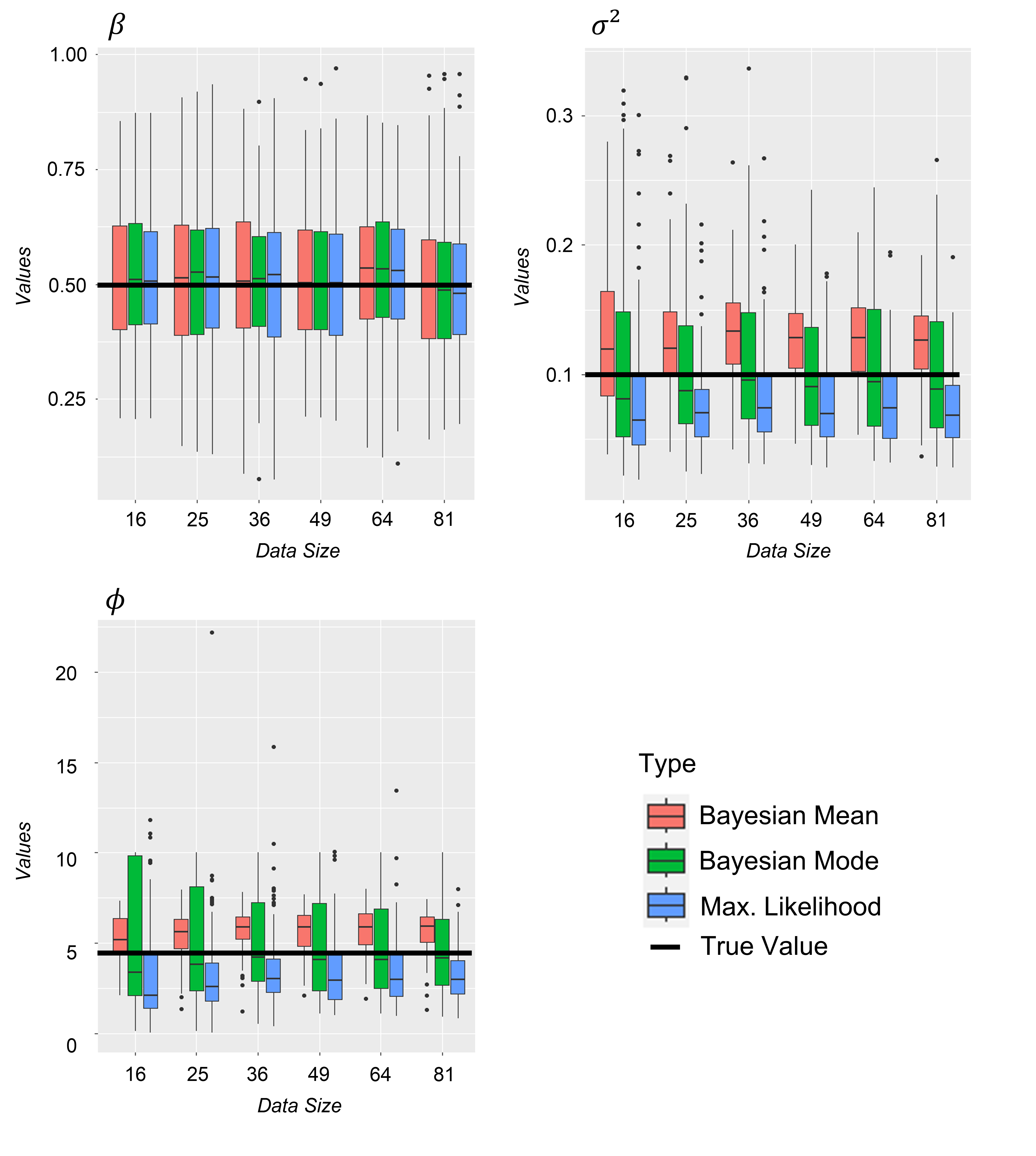}
\captionof{figure}{Boxplots of maximum likelihood estimated parameters, means and modes of the posterior distribution obtained with Bayesian approach, as a function of dataset size $n$. The results are obtained from datasets of $100$ independent draws of Gaussian processes.}
\label{fig:11}
\end{center}
\end{figure}

To get a better understanding of both kriging models, we choose to compare parameter estimation of both methods w.r.t. the number of observations. To do so, we consider a protocol similar to the one given in Sect. \ref{sec:4.1}. 
We use $100$ simulated datasets for a variable number of observations (here between $n=16$ and $n=81$ observations). For each of these simulated datasets, we compute on one side the estimated parameters for ordinary kriging by maximum likelihood method. On the other side, for Bayesian kriging, the a posteriori distribution of parameters is simulated relying on Bayes' theorem and Markov chain Monte Carlo methods. The covariance model and ``true'' parameter values used to simulated the datasets are identical to those presented in Sect. \ref{sec:4.1}. As the Bayesian approach produces an a posteriori distribution, we have chosen to represent the results obtained by considering both the mode and the mean of this distribution. The results are summarized by boxplots in Fig. \ref{fig:11}. 

The estimate of $\beta$ by both approaches remains close to the true mean, except in the case of $n=64$ observations where the results are slightly worse. The results between the maximum likelihood and Bayesian estimates (considering mean or the mode of the posterior distribution) are similar. In contrast, regarding the variance and correlation length, we observe that the methods produce significant differences. The maximum likelihood underestimates the variance, while the mean of the posterior distribution obtained with Bayesian kriging overestimates it. Considering the mode of this posterior distribution leads to better results on average, but at the cost of greater variability. The same observations can be made for the correlation length $\phi$.

\subsection{Posterior distribution of $\phi$ on simulated datasets of different sizes}

Another advantage of Bayesian kriging is the estimation of the posterior distribution for each parameter. This estimation allows to quantify more precisely the uncertainty associated with parameters estimation. For example, the posterior distribution of the correlation length $\phi$ is estimated with $1000$ samples \citep{diggle_bayesian_2002} and with the help of the discretization of $\phi$'s prior. A posterior density $d_\phi$ is then approximated using a Gaussian kernel.

\begin{figure}[!ht]
\begin{center}
\includegraphics[scale=0.29]{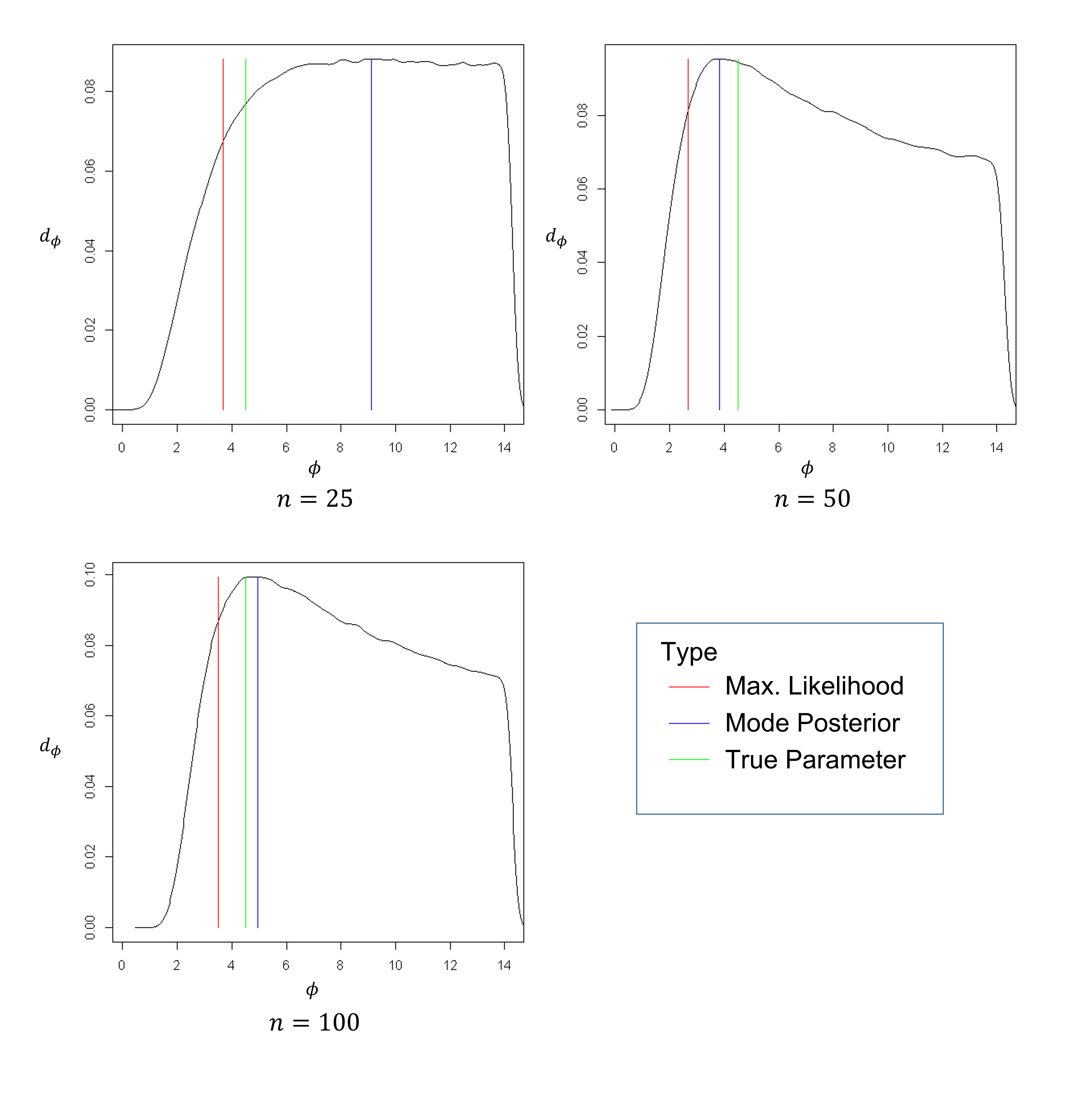}
\captionof{figure}{Posterior distribution of the correlation length for a random dataset of n = 25; 50 and 100 observations.}
\label{fig:12}
\end{center}
\end{figure}

The Fig. \ref{fig:12} illustrates the evolution of the posterior distribution as a function of the size of the data set and how the prior information has been updated by the addition of observations. When the number of available observations is small, the posterior distribution remains similar to the prior distribution (in this case a uniform prior): the observations provide little new information. On the other hand, as the number of observations increases, the mode of the distribution becomes closer to the true parameter.

\vspace{0.5cm}
\noindent {\bf Acknowledgements} 
We warmly thank C\'eline Helbert and Delphine Blanke for useful discussions.
We are also grateful to the associate editor and two anonymous referees for their very helpful comments on this paper.

\vspace{0.5cm}
\noindent {\bf Data Availability Statement}
The R codes of the numerical tests can be found in the GitLab repository: \url{https://gitlab.com/biooss/r-code-for-wieskotten-et-al-2023-paper}.

\bigskip
\bibliographystyle{MG}       
{\footnotesize
\bibliography{krigebayes}}   

\end{document}